\documentstyle[preprint,aps,epic,eepic,axodraw]{revtex}
\begin{document}
\preprint{\vbox{
\hbox{IFP-801-UNC}
\hbox{hep-ph/0110300}\hbox{\today} }}

\newcommand{\beq}{\begin{equation}}
\newcommand{\eeq}{\end{equation}}
\def\bea{\begin{eqnarray}}
\def\eea{\end{eqnarray}}
\def\nn{\nonumber}

\draft
\title{Zee Model Confronts SNO Data}
\author{{\bf Paul H. Frampton, Myoung C. Oh and Tadashi Yoshikawa}}
\address{Institute of Field Physics, Department of Physics and Astronomy,\\
University of North Carolina, Chapel Hill, NC 27599-3255.}

\maketitle

\begin{abstract}
We reexamine the solution of the minimal 
Zee model by comparing with the data of 
the SNO experiment, and conclude that the model is strongly disfavored but not yet 
excluded by the observations.  Two extensions of the Zee model are briefly discussed 
both of which introduce additional freedom and 
can accommodate the data.  
\end{abstract}


\bigskip

\newpage

\bigskip

\section{Introduction}
To understand why neutrino masses are non-zero is one of the 
most important subjects in particle physics. 
Assuming only left-handed neutrinos
$\nu_{aL}$ ($a = e, \mu, \tau$)
the minimum standard model predicts $m(\nu_a) = 0$
since adding a bare Majorana
mass term $M_{ab} \overline{\nu_{a L}^{c}} \nu_{bL}$
with $M_{ab} = M_{ba}$ for these neutrinos violates
gauge invariance (because
of the concomitant $\overline{e_{a L}^{c}} e_{b L}$ - type terms) 
and hence renormalizability.
Thus, further states must be added to accommodate the neutrino
masses, and we look
for the greatest economy and simplicity in doing this.
For a review of the theory, see \cite{Barr}.

\section{Minimal Zee model} 
The Zee model\cite{Zee} is one of the most
economical possible scenarios. In the Zee model,
the Majorana neutrino masses are generated by a one loop diagram. 
The origin of the smallness of the masses come from this feature. 
Hence one of the present authors has discussed the comparison 
with the neutrino 
experimental data in Ref.\cite{FG}. And several other authors have also 
discussed it and shown similar results \cite{Ja,Koide}.  
The solution of the neutrino mixing is bimaximal ( $\theta_1=\theta_3=\pi/4, 
\theta_2=0 $ with $\theta_i$ as defined below). 
This agrees with the atmospheric neutrino data. 
For the global analysis of solar neutrino data\cite{BGP}, 
this solution corresponds to 
the large-angle MSW (LMA) solution  
or the just-so vacuum oscillations (VAC). 
However the result from the Zee model 
does not agree so well with the LMA solution from the recent analysis
\cite{FG} and 
the recent data of the super-Kamiokande\cite{SK} disfavors the VAC solution.
So we need to reconsider the possibility of the LMA solution 
in the Zee model.  

Recently the SNO group announced their experimental data of solar 
neutrino fluxes from ${^8B }$ decay measured by the charged current reaction
rate, which is $\phi^{CC}(\nu_e) = 1.75\pm0.07^{+0.12}_{-0.11}\pm0.05 \times
10^6 cm^{-2}s^{-1}$. By combining this with the data from 
the Super-Kamiokande and using the values of the total flux expected in 
the Standard Solar Model (SSM)\cite{SSM}, 
the survival probability of $\nu_e $ was reported 
in ref.\cite{SNO}. It is 
\bea
P(\nu_e\rightarrow \nu_e) = 
    \frac{\phi^{CC}}{\phi_{SSM}} = 0.347\pm0.029^{+0.056}_{-0.069}
\label{SNO1}
\eea
where the first error is from SNO and the second from the 
SSM(Solar Standard model) theoretical error. Using a combination
 of $\phi^{CC}(\nu_e)$
and $\phi^{ES}(\nu_x)$ from SNO and Super-Kamiokande leads to
the estimate\cite{SNO} 
\bea
P(\nu_e\rightarrow \nu_e) = 0.322 \pm 0.076. 
\label{SNO2}
\eea
This suggests the survival probability $P(\nu_e\rightarrow \nu_e)$ 
is nearer to
$1/3$ than $1/2$ and this is new information for us to analyze. 

In this paper, we rediscuss the compatibility of the answer from Zee model 
with the LMA solution\cite{Barger} by comparing with the recent SNO data.

Under the Zee anzatz, the neutrino mass matrix 
in the flavor basis $(e,\mu,\tau)$ is
\bea
{\cal M } = \left( \begin{array}{ccc} 
                    0 & M_{e\mu} & M_{e\tau} \\
                    M_{e\mu} & 0 & M_{\mu\tau} \\
                    M_{e\tau}& M_{\mu\tau}& 0 \\
                   \end{array} \right) 
          = U \left( \begin{array}{ccc} 
                    m_1 & 0 & 0 \\
                    0 & m_2 & 0 \\
                    0& 0 & m_3 \\
                   \end{array} \right) U^\dagger 
\eea
where $m_1,m_2,m_3$ are the eigenvalues of ${\cal M }$ and $U$ is the unitary
matrix to diagonalize it.  ${\cal M }$ is real, traceless and symmetric. 
{}From the tracelessness condition, 
\bea
m_1 + m_2 + m_3 =0.
\eea
This condition is a strong constraint. 
The mass pattern of exact solutions which satisfy the atmospheric 
neutrino data 
is \cite{FG} 
\bea
m_1=-m_2, ~~~ m_3=0, 
\eea
and the ratio between the two neutrino squared-mass differences is 
$r=\Delta_s/\Delta_a = |m_1^2-m_2^2|/|m_1^2-m_3^2|=0$, where subscripts $s,a$
refer to solar, atmospheric respectively.
We will examine $r>0$ later.

With this situation, the allowed mixing matrix is the bimaximal one with
$\theta_1=\pi/4,\theta_2=0$ and $\theta_3=\pi/4$, where the definition 
of the mixing angle is\footnote{There is no CP violation in Zee model.}.
\bea
U= \left( \begin{array}{ccc}
       c_2 c_3 & c_2 s_3 & s_2 \\
       -c_1 s_3 - s_1 s_2 c_3 & c_1 c_3 - s_1 s_2 s_3 & s_1 c_2 \\
      s_1 s_3 - c_1 s_2 c_3 & - s_1 c_3 - c_1 s_2 s_3 & c_1 c_2 \\
          \end{array} \right),
\eea
with $s_i$ and $c_i$ standing for sines and cosines of $\theta_i$ and
the bimaximal mixing matrix is  
\bea
U=\left( \begin{array}{ccc}
         \frac{1}{\sqrt{2}} & \frac{1}{\sqrt{2}} & 0 \\
         -\frac{1}{2} & ~~\frac{1}{2} & \frac{1}{\sqrt{2}} \\
         ~~\frac{1}{2}   & -\frac{1}{2} & \frac{1}{\sqrt{2}} \\
         \end{array} \right).
\label{bimix}
\eea
To discuss the neutrino flux from the sun, we have to solve 
the neutrino propagation equation in the matter as follows: 
\bea
i\frac{d}{dt}\left( \begin{array}{c}
       \nu_e \\
       \nu_\mu \\
       \nu_\tau \\
       \end{array} \right) = \frac{1}{2E} M^2 \left( \begin{array}{c}
                                                 \nu_e \\
                                                 \nu_\mu \\
                                                 \nu_\tau \\
                                                 \end{array} \right)
       = \frac{1}{2E}\left[ U \left( \begin{array}{ccc}
                                m_1^2 & &  \\
                                      & m_2^2 &  \\
                                      & & m_3^2  \\
                                \end{array} \right) U^\dagger +
                         \left( \begin{array}{ccc}
                                A & ~~& ~ \\
                                      & 0  & ~ \\
                                      &~~ & 0 \\
                                \end{array} \right) \right]
               \left( \begin{array}{c}
       \nu_e \\
       \nu_\mu \\
       \nu_\tau \\
       \end{array} \right)
\label{peq}
\eea 
where $ A = 2 \sqrt{2} G_F N_e E $, $N_e$ is the density of electron 
neutrino in the sun, $E$ is the energy of the neutrino.
On the condition of eq.(5), $|m_1|=|m_2|$ and $m_3=0$, 
\bea
M^2= \left( \begin{array}{ccc}
            m_1^2  +A & 0 & 0 \\
            0 & \frac{1}{2}m_1^2 & -\frac{1}{2}m_1^2 \\
            0 & -\frac{1}{2}m_1^2 & \frac{1}{2}m_1^2 \\
             \end{array} \right) 
\eea
Then, the rotation among the weak eigenstates and the mass eigenstates in the
center of the Sun 
$(\nu_1^m,\nu_2^m,\nu_3^m) $ 
can be expressed as follows:
\bea
\left( \begin{array}{c}
       \nu_e \\
       \nu_\mu \\
       \nu_\tau \\
       \end{array} \right) = U_m \left( \begin{array}{c}
                                  \nu_1^m \\
                                  \nu_2^m \\
                                  \nu_3^m \\
                                  \end{array} \right) 
                           = \left( \begin{array}{ccc}
            0 & 1 & 0 \\
            \frac{1}{\sqrt{2}} & 0 & \frac{1}{\sqrt{2}} \\
           -\frac{1}{\sqrt{2}} & 0 & \frac{1}{\sqrt{2}} \\
             \end{array} \right) \left( \begin{array}{c}
                                  \nu_1^m \\
                                  \nu_2^m \\
                                  \nu_3^m \\
                                  \end{array} \right), 
\eea
where we have taken the limit of large electron neutrino 
density, ($A \rightarrow \infty$). 
At $t=0$ an 
electron neutrino is produced in the sun and it is composed mainly
of the state $\nu_2^m $ in the hierarchy
\footnote{As we show later, 
in the case $m_3 \neq 0$, the mass hierarchy we need is $m_2 > m_1 \gg m_3$.
By the matter effect, the hierarchy in dense matter is 
$m_2 \gg m_1 \gg m_3$. So we can define that the electron neutrino 
at the production point is composed mainly of the state $\nu_2$.}
we are considering in this work. 
\bea
|\nu_e(0)> = |\nu_2^m>.
\eea
The time evolution of this state to time t is
\bea
|\nu_e(t)> = e^{i\int^t_0\frac{\lambda_2}{2E}dt }|\nu_2^m>
\eea
where $\lambda_2$ is the eigenvalue of $M^2$ for $\nu_2^m$ state. 
Since the neutrino is measured on the earth, the state of $\nu_e $ is 
expressed by the mixing Eq.(\ref{bimix}). The amplitude is 
\bea
<\nu_e|\nu_e(t)> = \frac{1}{\sqrt{2}} e^{i\int\frac{\lambda_2}{2E}dt }.
\eea
Hence the survival probability is
\bea
P(\nu_e\rightarrow \nu_e) = |<\nu_e|\nu_e(t)>|^2 = \frac{1}{2}. 
\eea
This is the result from the exact Zee anzatz with $r = 0$ 
and is significantly disfavored by SNO data.  
We are led to consider a more realistic case, 
$r \neq 0$ and 
$|m_1| \neq |m_2|, m_3 \neq 0$.
For the more realistic case, we rewrite the mixing matrix of 
Eq.(\ref{bimix}) by using
the following parameters which show the discrepancy from exact bimaximal
mixing. 
\bea
c_1 &=& \cos(\frac{\pi}{4}-\xi_1 ) \sim \frac{1}{\sqrt{2}} (1+\varepsilon_1
                                     - \frac{1}{2}\varepsilon_1^2 ), 
\label{c1}\\
s_1 &=& \sin(\frac{\pi}{4}-\xi_1 ) \sim \frac{1}{\sqrt{2}} (1-\varepsilon_1
                                     - \frac{1}{2}\varepsilon_1^2 ), \\
c_2 &=& \cos(\xi_2) \sim (1
                                     - \frac{1}{2}\varepsilon_2^2 ), \\
s_2 &=& \sin(\xi_2 ) \sim \varepsilon_2, \\
c_3 &=& \cos(\frac{\pi}{4}-\xi_3 ) \sim \frac{1}{\sqrt{2}} (1+\varepsilon_3
                                     - \frac{1}{2}\varepsilon_3^2 ), \\
s_3 &=& \sin(\frac{\pi}{4}-\xi_3 ) \sim \frac{1}{\sqrt{2}} (1-\varepsilon_3
                                     - \frac{1}{2}\varepsilon_3^2 ), 
\label{s3}
\eea
where we neglected $\varepsilon_x^4$ and $\xi_x$ is the 
difference of the angle from the bimaximal case and
$\varepsilon_x \equiv \sin\xi_x $. 
By using this expansion, we can find the following relations 
up to $O(\varepsilon^2)$ 
among the parameters from the conditions that the diagonal element 
of ${\cal M}$ are zero,
\bea
m_1+m_2 &=& - 2 \varepsilon_3 (m_1-m_2), \\
\varepsilon_2 &=& 8\varepsilon_1\varepsilon_3. 
\eea
By these relation and the tracelessness condition $m_1+m_2+m_3=0$, we find 
the relation between $r=\Delta_s /\Delta_a$ and $\varepsilon_3 $ to be
\bea
r= \frac{|m_1^2-m_2^2|}{|m_1^2-m_3^2|}
      =\frac{ 8 \varepsilon_3}{1-4\varepsilon_3-12\varepsilon_3^2}.
\eea
The behavior of $r$ versus $\varepsilon_3 $ is shown in Fig.1. 
The experimental data suggest that the ratio $r$ satisfies 
$r \leq 0.1$ at 90\% confidence level.
So to satisfy this upper bound on $r$, $\varepsilon_3 $ cannot take too large 
values. From Fig.1, we find the magnitude of $\varepsilon_3 $ is smaller 
than about $0.015$. 

{}From the neutrino propagation equation, Eq.(\ref{peq}), 
after replacing $m_3$ by $m_3^2 = - 2 \epsilon_3 \Delta $, 
which comes from Eq.(21) and the traceless condition,
we find 

\bea
M^2 = M_0^2 + M_1^2 + O(\varepsilon^2),
\eea
where
\bea
M_0^2 &=& \left( \begin{array}{ccc}
            \frac{1}{2}\Sigma +A & 0 & 0 \\
            0 & \frac{1}{4}\Sigma & -\frac{1}{4}\Sigma \\
            0 & -\frac{1}{4}\Sigma & \frac{1}{4}\Sigma \\
             \end{array} \right) 
\eea

\noindent and

\bea
M_1^2 &=& \left( \begin{array}{ccc}
             \varepsilon_3 \Delta 
           & -\frac{1}{2\sqrt{2}}
                (\Delta + \varepsilon_1 \Delta + \varepsilon_2 \Sigma )
           & \frac{1}{2\sqrt{2}}
                (\Delta - \varepsilon_1 \Delta - \varepsilon_2 \Sigma) \\
            -\frac{1}{2\sqrt{2}}
                (\Delta + \varepsilon_1 \Delta + \varepsilon_2 \Sigma )
           & \frac{1}{2}(\varepsilon_1 \Sigma - 3 \varepsilon_3 \Delta) 
           & -\frac{1}{2}\varepsilon_3 \Delta \\
            \frac{1}{2\sqrt{2}}
                (\Delta - \varepsilon_1 \Delta - \varepsilon_2 \Sigma ) 
           & -\frac{1}{2}\varepsilon_3 \Delta 
           & \frac{1}{2}(\varepsilon_1 \Sigma - 3 \varepsilon_3 \Delta) 
             \end{array} \right) 
\eea

\noindent in which $\Sigma = m_1^2 + m_2^2 $ and $\Delta=m_1^2-m_2^2 $. 
In the case of $\varepsilon_3 \neq 0$ and $\varepsilon_1=\varepsilon_2=0$
\footnote{The atmospheric neutrino data favored $\varepsilon_1 =0$ and from
Eq.(22) $\varepsilon_2 =0$ is also favored.}, 
the mixing matrix which diagonalizes $M^2$
in the matter is
\bea
U_m = \frac{1}{\sqrt{2+K^2}} \left( \begin{array}{ccc}
                       \sqrt{2} & K & 0 \\
             -\frac{K}{\sqrt{2}} & 1 & \frac{\sqrt{2+K^2}}{\sqrt{2}} \\
              \frac{K}{\sqrt{2}} & -1 & \frac{\sqrt{2+K^2}}{\sqrt{2}} \\
             \end{array} \right) 
\eea
where 
\bea
K=\sqrt{2}\frac{-A-2\varepsilon_3 \Delta 
       - \sqrt{(A+2\varepsilon_3 \Delta )^2 +\Delta^2 }}{\Delta }. 
\eea
This means that the mixing angle $\theta_3^m $ satisfies the following relation,
\bea
\cos\theta_3^m = \frac{\sqrt{2}}{\sqrt{2+K^2}},~~
\sin\theta_3^m = \frac{K}{\sqrt{2+K^2}}.
\eea


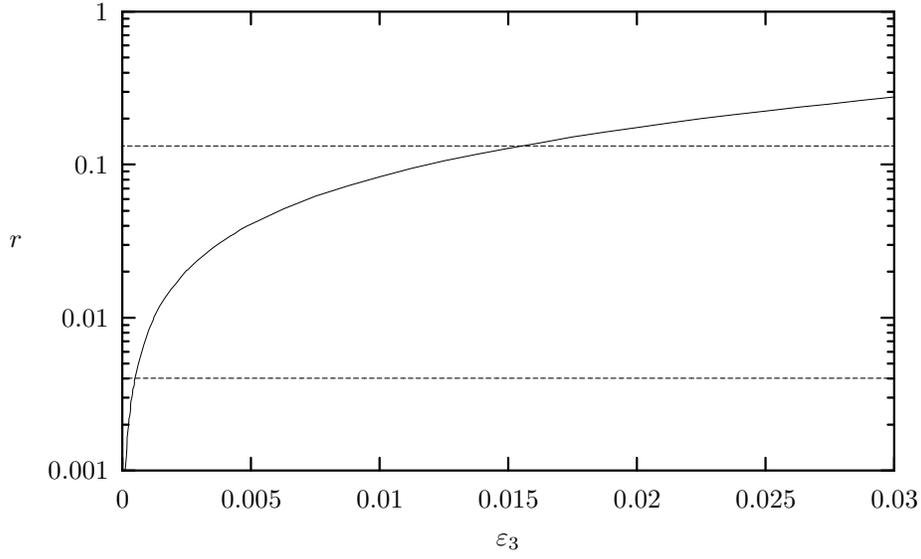
\begin{figure}[th]
\setlength{\unitlength}{0.240900pt}
\begin{picture}(1500,900)(0,0)
\footnotesize
\thicklines \path(221,135)(241,135)
\thicklines \path(1433,135)(1413,135)
\put(199,135){\makebox(0,0)[r]{0.001}}
\thicklines \path(221,207)(231,207)
\thicklines \path(1433,207)(1423,207)
\thicklines \path(221,250)(231,250)
\thicklines \path(1433,250)(1423,250)
\thicklines \path(221,280)(231,280)
\thicklines \path(1433,280)(1423,280)
\thicklines \path(221,303)(231,303)
\thicklines \path(1433,303)(1423,303)
\thicklines \path(221,322)(231,322)
\thicklines \path(1433,322)(1423,322)
\thicklines \path(221,338)(231,338)
\thicklines \path(1433,338)(1423,338)
\thicklines \path(221,352)(231,352)
\thicklines \path(1433,352)(1423,352)
\thicklines \path(221,364)(231,364)
\thicklines \path(1433,364)(1423,364)
\thicklines \path(221,375)(241,375)
\thicklines \path(1433,375)(1413,375)
\put(199,375){\makebox(0,0)[r]{0.01}}
\thicklines \path(221,448)(231,448)
\thicklines \path(1433,448)(1423,448)
\thicklines \path(221,490)(231,490)
\thicklines \path(1433,490)(1423,490)
\thicklines \path(221,520)(231,520)
\thicklines \path(1433,520)(1423,520)
\thicklines \path(221,543)(231,543)
\thicklines \path(1433,543)(1423,543)
\thicklines \path(221,562)(231,562)
\thicklines \path(1433,562)(1423,562)
\thicklines \path(221,578)(231,578)
\thicklines \path(1433,578)(1423,578)
\thicklines \path(221,592)(231,592)
\thicklines \path(1433,592)(1423,592)
\thicklines \path(221,605)(231,605)
\thicklines \path(1433,605)(1423,605)
\thicklines \path(221,616)(241,616)
\thicklines \path(1433,616)(1413,616)
\put(199,616){\makebox(0,0)[r]{0.1}}
\thicklines \path(221,688)(231,688)
\thicklines \path(1433,688)(1423,688)
\thicklines \path(221,730)(231,730)
\thicklines \path(1433,730)(1423,730)
\thicklines \path(221,760)(231,760)
\thicklines \path(1433,760)(1423,760)
\thicklines \path(221,784)(231,784)
\thicklines \path(1433,784)(1423,784)
\thicklines \path(221,803)(231,803)
\thicklines \path(1433,803)(1423,803)
\thicklines \path(221,819)(231,819)
\thicklines \path(1433,819)(1423,819)
\thicklines \path(221,833)(231,833)
\thicklines \path(1433,833)(1423,833)
\thicklines \path(221,845)(231,845)
\thicklines \path(1433,845)(1423,845)
\thicklines \path(221,856)(241,856)
\thicklines \path(1433,856)(1413,856)
\put(199,856){\makebox(0,0)[r]{1}}
\thicklines \path(221,135)(221,155)
\thicklines \path(221,856)(221,836)
\put(221,90){\makebox(0,0){0}}
\thicklines \path(423,135)(423,155)
\thicklines \path(423,856)(423,836)
\put(423,90){\makebox(0,0){0.005}}
\thicklines \path(625,135)(625,155)
\thicklines \path(625,856)(625,836)
\put(625,90){\makebox(0,0){0.01}}
\thicklines \path(827,135)(827,155)
\thicklines \path(827,856)(827,836)
\put(827,90){\makebox(0,0){0.015}}
\thicklines \path(1029,135)(1029,155)
\thicklines \path(1029,856)(1029,836)
\put(1029,90){\makebox(0,0){0.02}}
\thicklines \path(1231,135)(1231,155)
\thicklines \path(1231,856)(1231,836)
\put(1231,90){\makebox(0,0){0.025}}
\thicklines \path(1433,135)(1433,155)
\thicklines \path(1433,856)(1433,836)
\put(1433,90){\makebox(0,0){0.03}}
\thicklines \path(221,135)(1433,135)(1433,856)(221,856)(221,135)
\put(44,495){\makebox(0,0)[l]{\shortstack{$ r  $ }}}
\put(827,23){\makebox(0,0){$ \varepsilon_3 $}}
\thinlines \path(226,135)(227,152)(228,166)(229,178)(229,186)(230,199)(231,207)(232,216)(233,223)(234,231)(234,238)(235,244)(236,250)(237,255)(238,261)(238,263)(239,266)(240,271)(240,276)(247,304)(255,333)(263,357)(270,373)(271,376)(280,393)(289,407)(297,419)(306,430)(314,440)(322,449)(324,451)(330,457)(339,465)(347,472)(356,479)(364,485)(374,493)(381,497)(390,503)(398,508)(406,513)(415,518)(423,522)(474,546)(524,566)(576,583)(627,597)(676,610)(728,622)(778,632)(826,641)
\thinlines \path(826,641)(878,650)(927,659)(980,667)(1030,674)(1079,681)(1130,688)(1180,694)(1233,700)(1284,706)(1333,711)(1385,717)(1433,722)
\thinlines \dashline[-10]{18}(221,645)(221,645)(233,645)(245,645)(258,645)(270,645)(282,645)(294,645)(307,645)(319,645)(331,645)(343,645)(356,645)(368,645)(380,645)(392,645)(405,645)(417,645)(429,645)(441,645)(454,645)(466,645)(478,645)(490,645)(503,645)(515,645)(527,645)(539,645)(552,645)(564,645)(576,645)(588,645)(601,645)(613,645)(625,645)(637,645)(649,645)(662,645)(674,645)(686,645)(698,645)(711,645)(723,645)(735,645)(747,645)(760,645)(772,645)(784,645)(796,645)(809,645)(821,645)
\thinlines \dashline[-10]{18}(821,645)(833,645)(845,645)(858,645)(870,645)(882,645)(894,645)(907,645)(919,645)(931,645)(943,645)(956,645)(968,645)(980,645)(992,645)(1005,645)(1017,645)(1029,645)(1041,645)(1053,645)(1066,645)(1078,645)(1090,645)(1102,645)(1115,645)(1127,645)(1139,645)(1151,645)(1164,645)(1176,645)(1188,645)(1200,645)(1213,645)(1225,645)(1237,645)(1249,645)(1262,645)(1274,645)(1286,645)(1298,645)(1311,645)(1323,645)(1335,645)(1347,645)(1360,645)(1372,645)(1384,645)(1396,645)(1409,645)(1421,645)(1433,645)
\thinlines \dashline[-10]{18}(221,280)(221,280)(233,280)(245,280)(258,280)(270,280)(282,280)(294,280)(307,280)(319,280)(331,280)(343,280)(356,280)(368,280)(380,280)(392,280)(405,280)(417,280)(429,280)(441,280)(454,280)(466,280)(478,280)(490,280)(503,280)(515,280)(527,280)(539,280)(552,280)(564,280)(576,280)(588,280)(601,280)(613,280)(625,280)(637,280)(649,280)(662,280)(674,280)(686,280)(698,280)(711,280)(723,280)(735,280)(747,280)(760,280)(772,280)(784,280)(796,280)(809,280)(821,280)
\thinlines \dashline[-10]{18}(821,280)(833,280)(845,280)(858,280)(870,280)(882,280)(894,280)(907,280)(919,280)(931,280)(943,280)(956,280)(968,280)(980,280)(992,280)(1005,280)(1017,280)(1029,280)(1041,280)(1053,280)(1066,280)(1078,280)(1090,280)(1102,280)(1115,280)(1127,280)(1139,280)(1151,280)(1164,280)(1176,280)(1188,280)(1200,280)(1213,280)(1225,280)(1237,280)(1249,280)(1262,280)(1274,280)(1286,280)(1298,280)(1311,280)(1323,280)(1335,280)(1347,280)(1360,280)(1372,280)(1384,280)(1396,280)(1409,280)(1421,280)(1433,280)
\end{picture}
\caption{ $r$
as a function of $\varepsilon_3 $.}
\end{figure}

\bigskip
\bigskip
\bigskip

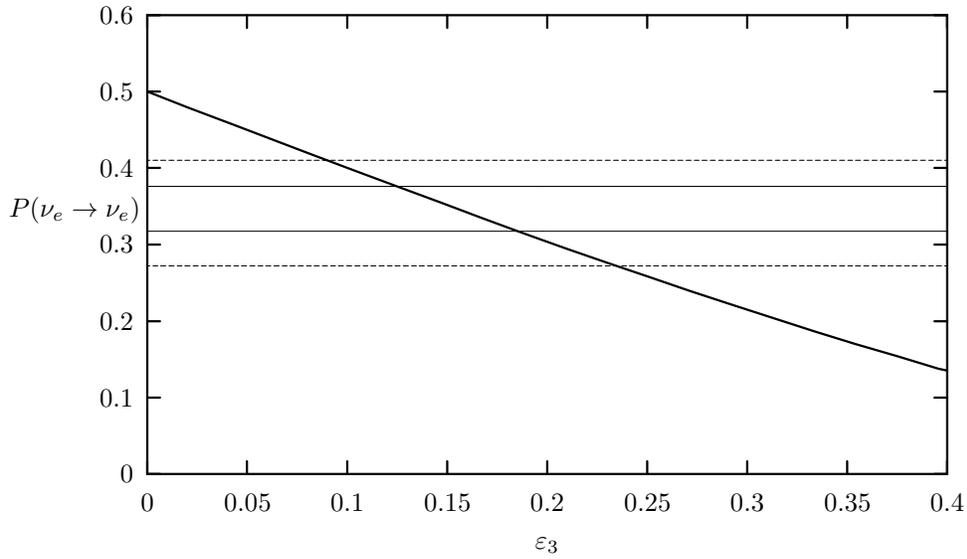
\begin{figure}[th]
\setlength{\unitlength}{0.240900pt}
\begin{picture}(1500,900)(0,0)
\footnotesize
\thicklines \path(177,135)(197,135)
\thicklines \path(1433,135)(1413,135)
\put(155,135){\makebox(0,0)[r]{0}}
\thicklines \path(177,255)(197,255)
\thicklines \path(1433,255)(1413,255)
\put(155,255){\makebox(0,0)[r]{0.1}}
\thicklines \path(177,375)(197,375)
\thicklines \path(1433,375)(1413,375)
\put(155,375){\makebox(0,0)[r]{0.2}}
\thicklines \path(177,496)(197,496)
\thicklines \path(1433,496)(1413,496)
\put(155,496){\makebox(0,0)[r]{0.3}}
\thicklines \path(177,616)(197,616)
\thicklines \path(1433,616)(1413,616)
\put(155,616){\makebox(0,0)[r]{0.4}}
\thicklines \path(177,736)(197,736)
\thicklines \path(1433,736)(1413,736)
\put(155,736){\makebox(0,0)[r]{0.5}}
\thicklines \path(177,856)(197,856)
\thicklines \path(1433,856)(1413,856)
\put(155,856){\makebox(0,0)[r]{0.6}}
\thicklines \path(177,135)(177,155)
\thicklines \path(177,856)(177,836)
\put(177,90){\makebox(0,0){0}}
\thicklines \path(334,135)(334,155)
\thicklines \path(334,856)(334,836)
\put(334,90){\makebox(0,0){0.05}}
\thicklines \path(491,135)(491,155)
\thicklines \path(491,856)(491,836)
\put(491,90){\makebox(0,0){0.1}}
\thicklines \path(648,135)(648,155)
\thicklines \path(648,856)(648,836)
\put(648,90){\makebox(0,0){0.15}}
\thicklines \path(805,135)(805,155)
\thicklines \path(805,856)(805,836)
\put(805,90){\makebox(0,0){0.2}}
\thicklines \path(962,135)(962,155)
\thicklines \path(962,856)(962,836)
\put(962,90){\makebox(0,0){0.25}}
\thicklines \path(1119,135)(1119,155)
\thicklines \path(1119,856)(1119,836)
\put(1119,90){\makebox(0,0){0.3}}
\thicklines \path(1276,135)(1276,155)
\thicklines \path(1276,856)(1276,836)
\put(1276,90){\makebox(0,0){0.35}}
\thicklines \path(1433,135)(1433,155)
\thicklines \path(1433,856)(1433,836)
\put(1433,90){\makebox(0,0){0.4}}
\thicklines \path(177,135)(1433,135)(1433,856)(177,856)(177,135)
\put(-40,550){\makebox(0,0)[l]{\shortstack{$ P(\nu_e \to \nu_e) $ }}}
\put(805,23){\makebox(0,0){$ \varepsilon_3 $}}
\thicklines \path(177,736)(177,736)(241,711)(310,685)(375,660)(438,636)(505,611)(569,587)(638,562)(703,537)(767,514)(834,490)(899,467)(961,446)(1028,423)(1092,402)(1160,380)(1225,359)(1288,340)(1355,320)(1419,301)(1433,298)
\thinlines \path(177,587)(177,587)(190,587)(202,587)(215,587)(228,587)(240,587)(253,587)(266,587)(278,587)(291,587)(304,587)(317,587)(329,587)(342,587)(355,587)(367,587)(380,587)(393,587)(405,587)(418,587)(431,587)(443,587)(456,587)(469,587)(481,587)(494,587)(507,587)(520,587)(532,587)(545,587)(558,587)(570,587)(583,587)(596,587)(608,587)(621,587)(634,587)(646,587)(659,587)(672,587)(684,587)(697,587)(710,587)(723,587)(735,587)(748,587)(761,587)(773,587)(786,587)(799,587)
\thinlines \path(799,587)(811,587)(824,587)(837,587)(849,587)(862,587)(875,587)(887,587)(900,587)(913,587)(926,587)(938,587)(951,587)(964,587)(976,587)(989,587)(1002,587)(1014,587)(1027,587)(1040,587)(1052,587)(1065,587)(1078,587)(1090,587)(1103,587)(1116,587)(1129,587)(1141,587)(1154,587)(1167,587)(1179,587)(1192,587)(1205,587)(1217,587)(1230,587)(1243,587)(1255,587)(1268,587)(1281,587)(1293,587)(1306,587)(1319,587)(1332,587)(1344,587)(1357,587)(1370,587)(1382,587)(1395,587)(1408,587)(1420,587)(1433,587)
\thinlines \path(177,517)(177,517)(190,517)(202,517)(215,517)(228,517)(240,517)(253,517)(266,517)(278,517)(291,517)(304,517)(317,517)(329,517)(342,517)(355,517)(367,517)(380,517)(393,517)(405,517)(418,517)(431,517)(443,517)(456,517)(469,517)(481,517)(494,517)(507,517)(520,517)(532,517)(545,517)(558,517)(570,517)(583,517)(596,517)(608,517)(621,517)(634,517)(646,517)(659,517)(672,517)(684,517)(697,517)(710,517)(723,517)(735,517)(748,517)(761,517)(773,517)(786,517)(799,517)
\thinlines \path(799,517)(811,517)(824,517)(837,517)(849,517)(862,517)(875,517)(887,517)(900,517)(913,517)(926,517)(938,517)(951,517)(964,517)(976,517)(989,517)(1002,517)(1014,517)(1027,517)(1040,517)(1052,517)(1065,517)(1078,517)(1090,517)(1103,517)(1116,517)(1129,517)(1141,517)(1154,517)(1167,517)(1179,517)(1192,517)(1205,517)(1217,517)(1230,517)(1243,517)(1255,517)(1268,517)(1281,517)(1293,517)(1306,517)(1319,517)(1332,517)(1344,517)(1357,517)(1370,517)(1382,517)(1395,517)(1408,517)(1420,517)(1433,517)
\thinlines \dashline[-10]{18}(177,628)(177,628)(190,628)(202,628)(215,628)(228,628)(240,628)(253,628)(266,628)(278,628)(291,628)(304,628)(317,628)(329,628)(342,628)(355,628)(367,628)(380,628)(393,628)(405,628)(418,628)(431,628)(443,628)(456,628)(469,628)(481,628)(494,628)(507,628)(520,628)(532,628)(545,628)(558,628)(570,628)(583,628)(596,628)(608,628)(621,628)(634,628)(646,628)(659,628)(672,628)(684,628)(697,628)(710,628)(723,628)(735,628)(748,628)(761,628)(773,628)(786,628)(799,628)
\thinlines \dashline[-10]{18}(799,628)(811,628)(824,628)(837,628)(849,628)(862,628)(875,628)(887,628)(900,628)(913,628)(926,628)(938,628)(951,628)(964,628)(976,628)(989,628)(1002,628)(1014,628)(1027,628)(1040,628)(1052,628)(1065,628)(1078,628)(1090,628)(1103,628)(1116,628)(1129,628)(1141,628)(1154,628)(1167,628)(1179,628)(1192,628)(1205,628)(1217,628)(1230,628)(1243,628)(1255,628)(1268,628)(1281,628)(1293,628)(1306,628)(1319,628)(1332,628)(1344,628)(1357,628)(1370,628)(1382,628)(1395,628)(1408,628)(1420,628)(1433,628)
\thinlines \dashline[-10]{18}(177,462)(177,462)(190,462)(202,462)
(215,462)(228,462)(240,462)
(253,462)(266,462)(278,462)(291,462)(304,462)(317,462)(329,462)
(342,462)(355,462)(367,462)(380,462)(393,462)(405,462)(418,462)
(431,462)(443,462)
(456,462)(469,462)(481,462)(494,462)(507,462)(520,462)(532,462)(545,462)
(558,462)(570,462)(583,462)(596,462)(608,462)(621,462)(634,462)(646,462)
(659,462)(672,462)(684,462)(697,462)(710,462)(723,462)(735,462)(748,462)
(761,462)(773,462)(786,462)(799,462)
\thinlines \dashline[-10]{18}(799,462)(811,462)(824,462)(837,462)(849,462)(862,462)(875,462)
(887,462)(900,462)(913,462)(926,462)(938,462)(951,462)(964,462)(976,462)
(989,462)(1002,462)(1014,462)(1027,462)(1040,462)(1052,462)(1065,462)
(1078,462)(1090,462)(1103,462)(1116,462)(1129,462)(1141,462)(1154,462)
(1167,462)(1179,462)(1192,462)(1205,462)(1217,462)(1230,462)(1243,462)
(1255,462)(1268,462)(1281,462)(1293,462)(1306,462)(1319,462)(1332,462)
(1344,462)(1357,462)(1370,462)(1382,462)(1395,462)(1408,462)(1420,462)
(1433,462)

\end{picture}
\caption{The survival probability of the electron neutrino. The horizontal lines
show the bounds from the SNO experiment: the solid lines exclude the SSM theoretical error
while the dotted lines include it.
The diagonal line is from Eq. (\ref{30}) with 
$P_C$ set to zero. }
\end{figure}

\newpage

\noindent The production point of the neutrino from ${^8B}$ decay is 
in the central core of the Sun so that the electron density $A$ 
has a very large value. At the production point, $\sin\theta_3^m \sim 1 $. 
This means the electron neutrino at the production point 
is composed mainly 
of $\nu_2$ as for the earlier discussion on the case  
$m_3 = 0$. 
Then the survival probability is 
\bea
<P(\nu_e\rightarrow \nu_e)> 
     = \frac{1}{2} - \varepsilon_3 
                       (1-P_c) 
\label{30}
\eea
where $P_c$ is the hopping probability at the resonance point and almost zero
in the LMA region. The dependence of $\varepsilon_3$ on the 
survival probability is shown in Fig.2.  
The constraint from SNO experimental data is given in Eqs.(\ref{SNO1}) and 
(\ref{SNO2}).

To satisfy this condition, the parameter, $\varepsilon_3$, would need to
be larger than 
0.1. But then the value of $r$ becomes significantly greater than 1 
and this is very hard to accommodate within the data on $\Delta_a$ and $\Delta_s $. 
The data suggest $\Delta_a > 1.5 \times 10^{-3} eV^2$ \cite{SKatm}
while $\Delta _S < 2 \times 10^{-4} eV^2$\cite{KS,krastev} at
90\% C.L.

We conclude that the minimal Zee model is strongly
disfavored but not yet fully excluded by the SNO/SuperKamiokande results.
For the minimal Zee model to survive, one would need $\Delta_S$ to get 
larger than $\Delta_a$.

\bigskip
\bigskip

\section{Extensions of the minimal Zee model}
Before looking at specific extended Zee models we consider 
adding diagonal elements in the neutrino mass matrix and relaxing 
the tracelessness condition to the more general 
\bea
m_1+m_2+m_3= \delta \times m_1 
\label{mtraceless}
\eea
where $\delta $ is the shift from $0$ and normalized by $m_1$. 
Then, we request the $M_{ee}=0$ in order to avoid the difficulty from 
double $\beta $ decay. We then find the constraints as follows:
\bea
m_1 + m_2 = - 2 \varepsilon_3 (m_1-m_2) ,
\label{mee=0} \\
m_3 = \frac{4 \varepsilon_3 + \delta 
    - 2 \varepsilon_3 \delta }{1-2\varepsilon_3}m_1. 
\eea
{}From these conditions, 
\bea
|m_1^2-m_2^2| &=& \frac{8 \varepsilon_3}{(1-2\varepsilon_3)^2 }m_1^2, \\
|m_1^2-m_3^2| &=& \frac{\{1 - \delta^2 - 4 \varepsilon_3(1 - \delta)^2 
                          - 4 \varepsilon_3^2 (3-\delta)(1-\delta) \} }
                   {(1-2\varepsilon_3)^2} m_1^2. 
\eea 
To satisfy the survival probability result requires that $\varepsilon_3 $ is
greater than about $0.1$, while $r$ must be smaller than $0.1$. However
this condition is not satisfied unless we keep the mass relation 
$m_1 > m_3$. The correlation between the ratio $r$ and $m_3/m_1$ when 
$\varepsilon_3 = 0.1,~0.15$ and $0.2$ is shown in Fig.3. 
In this case, to satisfy the conditions from experiments,
the mass hierarchy is $|m_3|\sim |\delta m_1| = 2 |M_{\mu\mu}| 
> |4 m_1| > |m_2| > |m_1| $ 
and the parameter $\delta $ has to be larger than about $4$ from Fig.3. 
So a very 
large deviation from the tracelessness condition is needed.

\begin{figure}
\setlength{\unitlength}{0.240900pt}
\begin{picture}(1500,900)(0,0)
\footnotesize
\thicklines \path(177,135)(197,135)
\thicklines \path(1433,135)(1413,135)
\put(155,135){\makebox(0,0)[r]{0}}
\thicklines \path(177,279)(197,279)
\thicklines \path(1433,279)(1413,279)
\put(155,279){\makebox(0,0)[r]{0.2}}
\thicklines \path(177,423)(197,423)
\thicklines \path(1433,423)(1413,423)
\put(155,423){\makebox(0,0)[r]{0.4}}
\thicklines \path(177,568)(197,568)
\thicklines \path(1433,568)(1413,568)
\put(155,568){\makebox(0,0)[r]{0.6}}
\thicklines \path(177,712)(197,712)
\thicklines \path(1433,712)(1413,712)
\put(155,712){\makebox(0,0)[r]{0.8}}
\thicklines \path(177,856)(197,856)
\thicklines \path(1433,856)(1413,856)
\put(155,856){\makebox(0,0)[r]{1}}
\thicklines \path(177,135)(177,155)
\thicklines \path(177,856)(177,836)
\put(177,90){\makebox(0,0){0}}
\thicklines \path(428,135)(428,155)
\thicklines \path(428,856)(428,836)
\put(428,90){\makebox(0,0){2}}
\thicklines \path(679,135)(679,155)
\thicklines \path(679,856)(679,836)
\put(679,90){\makebox(0,0){4}}
\thicklines \path(931,135)(931,155)
\thicklines \path(931,856)(931,836)
\put(931,90){\makebox(0,0){6}}
\thicklines \path(1182,135)(1182,155)
\thicklines \path(1182,856)(1182,836)
\put(1182,90){\makebox(0,0){8}}
\thicklines \path(1433,135)(1433,155)
\thicklines \path(1433,856)(1433,836)
\put(1433,90){\makebox(0,0){10}}
\thicklines \path(177,135)(1433,135)(1433,856)(177,856)(177,135)
\put(44,495){\makebox(0,0)[l]{\shortstack{$ r $ }}}
\put(805,23){\makebox(0,0){$ m_3/m_1 $}}
\thinlines \path(468,856)(473,786)(485,653)(491,596)(497,551)(511,474)(524,419)(536,381)(549,348)(575,299)(588,280)(602,263)(628,240)(655,222)(682,208)(710,197)(760,183)(786,177)(813,172)(868,165)(921,160)(971,156)(1025,153)(1077,150)(1127,148)(1180,147)(1231,145)(1286,144)(1338,143)(1389,142)(1433,142)
\thicklines \path(408,856)(410,801)(417,693)(424,616)(435,515)(442,473)(448,441)(461,387)(473,349)(486,316)(500,290)(514,270)(528,252)(555,228)(581,211)(609,198)(635,189)(659,182)(710,171)(738,166)(765,163)(815,158)(869,154)(921,151)(971,149)(1024,147)(1075,145)(1129,144)(1181,143)(1231,142)(1285,141)(1336,141)(1391,140)(1433,140)
\Thicklines \path(542,856)(542,850)(549,768)(555,704)(569,591)(582,514)(594,460)(606,414)(633,347)(646,322)(660,299)(686,268)(713,244)(741,225)(768,212)(818,194)(844,187)(872,180)(927,171)(980,165)(1030,160)(1085,156)(1137,153)(1187,151)(1240,149)(1291,147)(1346,146)(1399,145)(1433,144)
\end{picture}
\caption{The correlation between $r$ and $m_3$ }
\end{figure}
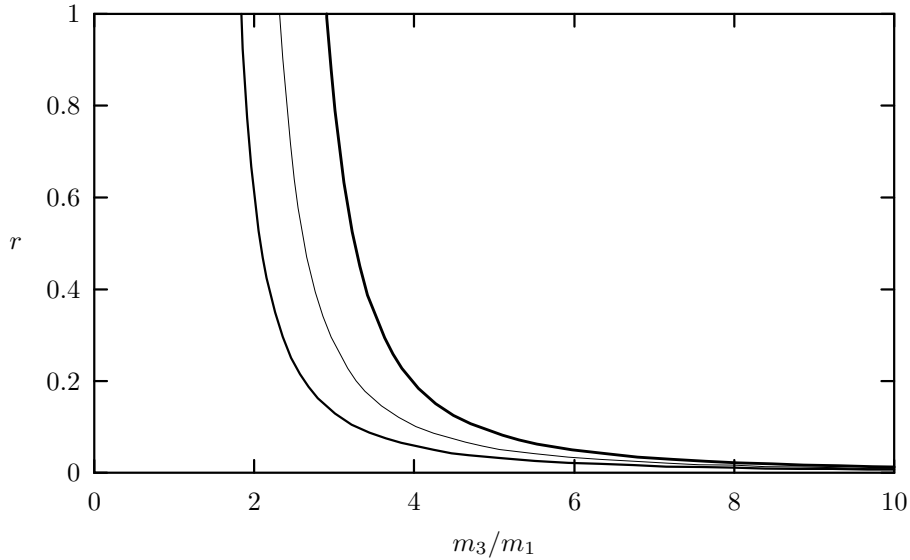

We consider the case that the all diagonal elements of ${\it M}$ 
are not zero. 
Here we parameterize the diagonal elements as follows,
\bea
M_{ee} &=& \delta_{ee} \times m_1, 
\label{mee=d}\\
M_{\mu\mu} &=& M_{\tau\tau } = \delta_{\mu\mu} \times m_1. 
\label{mmm=d}
\eea
Then, by the angles we defined in Eqs.(\ref{c1}) - (\ref{s3}), 
the relations Eqs.(\ref{mtraceless})
and (\ref{mee=0}) are changed to
\bea
m_1+m_2+m_3 &=& (\delta_{ee} + 2\delta_{\mu\mu} )m_1, \\
m_1+m_2 + 2 \varepsilon_3 (m_1-m_2) &=& 2 \delta_{ee} \times m_1. 
\eea
{}From these relations,
\bea
m_3=2 \varepsilon_3 (m_1 - m_2 ) -(\delta_{ee}-2 \delta_{\mu\mu}) m_1 .
\eea
And then,  
\bea
r=\frac{ 4 ( 1- \delta_{ee} ) (2 \varepsilon_3 - \delta_{ee} )}
       { (1-2\varepsilon_3)^2 - \{ (1-2\varepsilon_3)
                       (\delta_{ee} + 2\delta_{\mu\mu})  
                    - 2 ( 2 \varepsilon_3 - \delta_{ee} )\}^2 }. 
\label{ndiagR}
\eea
In Fig.4, $\Delta_S = r \Delta_a$, with $\Delta_a = 3 \times 10^{-3} eV^2$
is plotted versus $\delta_{ee}$ using Eq.(\ref{ndiagR}).
To reduce $r$, $\delta_{ee} $ should be near to $2\varepsilon_3$( or to 1), 
because the numerator of eq.(\ref{ndiagR}) has the factor of
$(2 \varepsilon_3 - \delta_{ee} )$. 
Hence, this condition will admit consistency with the LMA solution. However 
we have to note that $\delta_{ee} $ is constrained by neutrinoless 
double $\beta$ decay experiment. In this case ($\delta_{\mu\mu}=0$), 
the mass pattern is
\bea
m_1 \sim -m_2, ~~~~~ m_3 \sim 
                          M_{ee}\sim 2 \varepsilon_3 m_1 , 
\eea 
and $m_1 \sim O(10^{-2}) eV $. This is not yet excluded by neutrinoless 
double $\beta$ decay experiment. 

In this discussion, we found the element $M_{ee}$ is important to realize 
the LMA solution from the model based on the Zee model. 

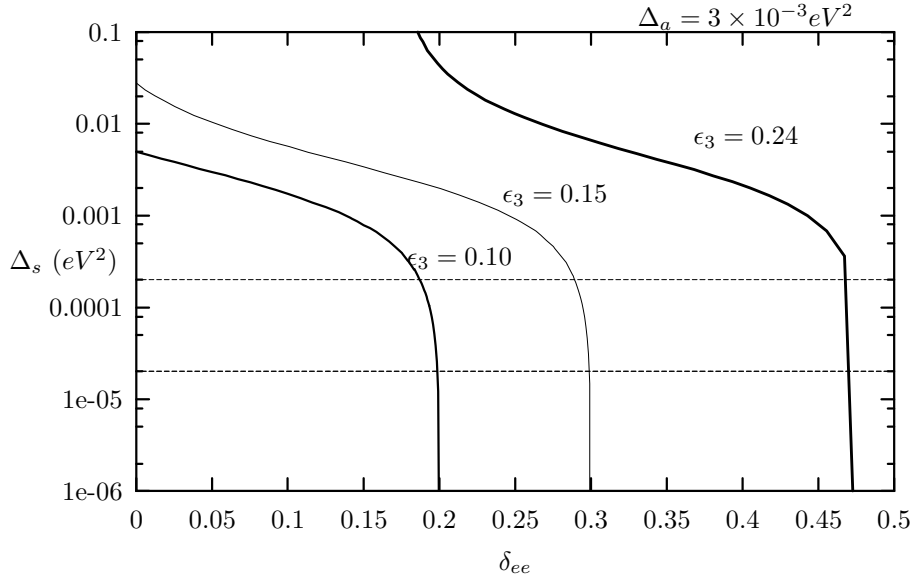
\begin{figure}
\setlength{\unitlength}{0.240900pt}
\begin{picture}(1500,900)(0,0)
\footnotesize
\thicklines \path(243,135)(263,135)
\thicklines \path(1433,135)(1413,135)
\put(221,135){\makebox(0,0)[r]{1e-06}}
\thicklines \path(243,178)(253,178)
\thicklines \path(1433,178)(1423,178)
\thicklines \path(243,236)(253,236)
\thicklines \path(1433,236)(1423,236)
\thicklines \path(243,265)(253,265)
\thicklines \path(1433,265)(1423,265)
\thicklines \path(243,279)(263,279)
\thicklines \path(1433,279)(1413,279)
\put(221,279){\makebox(0,0)[r]{1e-05}}
\thicklines \path(243,323)(253,323)
\thicklines \path(1433,323)(1423,323)
\thicklines \path(243,380)(253,380)
\thicklines \path(1433,380)(1423,380)
\thicklines \path(243,409)(253,409)
\thicklines \path(1433,409)(1423,409)
\thicklines \path(243,423)(263,423)
\thicklines \path(1433,423)(1413,423)
\put(221,423){\makebox(0,0)[r]{0.0001}}
\thicklines \path(243,467)(253,467)
\thicklines \path(1433,467)(1423,467)
\thicklines \path(243,524)(253,524)
\thicklines \path(1433,524)(1423,524)
\thicklines \path(243,554)(253,554)
\thicklines \path(1433,554)(1423,554)
\thicklines \path(243,568)(263,568)
\thicklines \path(1433,568)(1413,568)
\put(221,568){\makebox(0,0)[r]{0.001}}
\thicklines \path(243,611)(253,611)
\thicklines \path(1433,611)(1423,611)
\thicklines \path(243,668)(253,668)
\thicklines \path(1433,668)(1423,668)
\thicklines \path(243,698)(253,698)
\thicklines \path(1433,698)(1423,698)
\thicklines \path(243,712)(263,712)
\thicklines \path(1433,712)(1413,712)
\put(221,712){\makebox(0,0)[r]{0.01}}
\thicklines \path(243,755)(253,755)
\thicklines \path(1433,755)(1423,755)
\thicklines \path(243,813)(253,813)
\thicklines \path(1433,813)(1423,813)
\thicklines \path(243,842)(253,842)
\thicklines \path(1433,842)(1423,842)
\thicklines \path(243,856)(263,856)
\thicklines \path(1433,856)(1413,856)
\put(221,856){\makebox(0,0)[r]{0.1}}
\thicklines \path(243,135)(243,155)
\thicklines \path(243,856)(243,836)
\put(243,90){\makebox(0,0){0}}
\thicklines \path(362,135)(362,155)
\thicklines \path(362,856)(362,836)
\put(362,90){\makebox(0,0){0.05}}
\thicklines \path(481,135)(481,155)
\thicklines \path(481,856)(481,836)
\put(481,90){\makebox(0,0){0.1}}
\thicklines \path(600,135)(600,155)
\thicklines \path(600,856)(600,836)
\put(600,90){\makebox(0,0){0.15}}
\thicklines \path(719,135)(719,155)
\thicklines \path(719,856)(719,836)
\put(719,90){\makebox(0,0){0.2}}
\thicklines \path(838,135)(838,155)
\thicklines \path(838,856)(838,836)
\put(838,90){\makebox(0,0){0.25}}
\thicklines \path(957,135)(957,155)
\thicklines \path(957,856)(957,836)
\put(957,90){\makebox(0,0){0.3}}
\thicklines \path(1076,135)(1076,155)
\thicklines \path(1076,856)(1076,836)
\put(1076,90){\makebox(0,0){0.35}}
\thicklines \path(1195,135)(1195,155)
\thicklines \path(1195,856)(1195,836)
\put(1195,90){\makebox(0,0){0.4}}
\thicklines \path(1314,135)(1314,155)
\thicklines \path(1314,856)(1314,836)
\put(1314,90){\makebox(0,0){0.45}}
\thicklines \path(1433,135)(1433,155)
\thicklines \path(1433,856)(1433,836)
\put(1433,90){\makebox(0,0){0.5}}
\thicklines \path(243,135)(1433,135)(1433,856)(243,856)(243,135)
\put(44,495){\makebox(0,0)[l]{\shortstack{$ \Delta_s ~(eV^2) $ }}}
\put(838,23){\makebox(0,0){$ \delta_{ee} $}}
\put(750,500){\makebox(0,0){$ \epsilon_3=0.10 $}}
\put(900,600){\makebox(0,0){$ \epsilon_3=0.15 $}}
\put(1200,690){\makebox(0,0){$ \epsilon_3=0.24 $}}
\put(1200,880){\makebox(0,0){$ \Delta_a=3\times 10^{-3} eV^2 $}}
\thinlines \path(243,776)(243,776)(257,765)(272,756)(287,748)(304,739)(334,725)(363,714)(394,703)(424,693)(453,684)(484,676)(513,667)(541,660)(572,652)(601,644)(632,635)(661,627)(690,619)(721,610)(750,600)(781,589)(811,576)(840,562)(848,557)(858,551)(868,545)(871,543)(878,537)(887,529)(898,519)(900,516)(907,508)(918,494)(928,475)(929,473)(931,467)(933,463)(935,457)(937,451)(939,444)(941,437)(943,429)(945,419)(947,408)(949,393)(951,375)(953,350)(955,304)(955,135)
\thicklines \path(243,668)(243,668)(262,663)(283,657)(303,652)(322,647)(342,641)(362,636)(383,631)(403,625)(422,620)(442,614)(462,608)(481,602)(501,595)(520,588)(541,581)(561,573)(580,564)(581,563)(590,559)(600,553)(600,553)(610,548)(620,542)(620,542)(630,535)(640,527)(640,527)(649,519)(660,509)(661,508)(669,497)(671,495)(673,492)(675,489)(677,486)(679,483)(680,483)(680,483)(685,473)(687,469)(689,465)(691,461)(693,456)(695,451)(697,446)(699,440)(700,436)(701,433)(703,426)
\thicklines \path(703,426)(705,418)(707,408)(709,397)(711,382)(713,363)(715,338)(717,292)(718,135)
\Thicklines \path(685,856)(685,856)(689,847)(693,841)(697,833)(700,827)(708,817)(716,807)(724,798)(732,790)(748,777)(763,766)(778,757)(791,749)(821,735)(837,728)(852,722)(881,711)(912,700)(941,691)(970,682)(1000,673)(1029,665)(1060,656)(1090,648)(1119,640)(1149,630)(1179,621)(1207,611)(1237,599)(1266,586)(1297,568)(1327,544)(1355,504)(1368,135)
\thinlines \dashline[-10]{18}(243,323)(243,323)(255,323)(267,323)(279,323)(291,323)(303,323)(315,323)(327,323)(339,323)(351,323)(363,323)(375,323)(387,323)(399,323)(411,323)(423,323)(435,323)(447,323)(459,323)(471,323)(483,323)(495,323)(507,323)(519,323)(531,323)(544,323)(556,323)(568,323)(580,323)(592,323)(604,323)(616,323)(628,323)(640,323)(652,323)(664,323)(676,323)(688,323)(700,323)(712,323)(724,323)(736,323)(748,323)(760,323)(772,323)(784,323)(796,323)(808,323)(820,323)(832,323)
\thinlines \dashline[-10]{18}(832,323)(844,323)(856,323)(868,323)(880,323)(892,323)(904,323)(916,323)(928,323)(940,323)(952,323)(964,323)(976,323)(988,323)(1000,323)(1012,323)(1024,323)(1036,323)(1048,323)(1060,323)(1072,323)(1084,323)(1096,323)(1108,323)(1120,323)(1132,323)(1145,323)(1157,323)(1169,323)(1181,323)(1193,323)(1205,323)(1217,323)(1229,323)(1241,323)(1253,323)(1265,323)(1277,323)(1289,323)(1301,323)(1313,323)(1325,323)(1337,323)(1349,323)(1361,323)(1373,323)(1385,323)(1397,323)(1409,323)(1421,323)(1433,323)
\thinlines \dashline[-10]{18}(243,467)(243,467)(255,467)(267,467)(279,467)(291,467)(303,467)(315,467)(327,467)(339,467)(351,467)(363,467)(375,467)(387,467)(399,467)(411,467)(423,467)(435,467)(447,467)(459,467)(471,467)(483,467)(495,467)(507,467)(519,467)(531,467)(544,467)(556,467)(568,467)(580,467)(592,467)(604,467)(616,467)(628,467)(640,467)(652,467)(664,467)(676,467)(688,467)(700,467)(712,467)(724,467)(736,467)(748,467)(760,467)(772,467)(784,467)(796,467)(808,467)(820,467)(832,467)
\thinlines \dashline[-10]{18}(832,467)(844,467)(856,467)(868,467)(880,467)(892,467)(904,467)(916,467)(928,467)(940,467)(952,467)(964,467)(976,467)(988,467)(1000,467)(1012,467)(1024,467)(1036,467)(1048,467)(1060,467)(1072,467)(1084,467)(1096,467)(1108,467)(1120,467)(1132,467)(1145,467)(1157,467)(1169,467)(1181,467)(1193,467)(1205,467)(1217,467)(1229,467)(1241,467)(1253,467)(1265,467)(1277,467)(1289,467)(1301,467)(1313,467)(1325,467)(1337,467)(1349,467)(1361,467)(1373,467)(1385,467)(1397,467)(1409,467)(1421,467)(1433,467)
\thinlines \dashline[-10]{18}(243,323)(243,323)(255,323)(267,323)(279,323)(291,323)(303,323)(315,323)(327,323)(339,323)(351,323)(363,323)(375,323)(387,323)(399,323)(411,323)(423,323)(435,323)(447,323)(459,323)(471,323)(483,323)(495,323)(507,323)(519,323)(531,323)(544,323)(556,323)(568,323)(580,323)(592,323)(604,323)(616,323)(628,323)(640,323)(652,323)(664,323)(676,323)(688,323)(700,323)(712,323)(724,323)(736,323)(748,323)(760,323)(772,323)(784,323)(796,323)(808,323)(820,323)(832,323)
\thinlines \dashline[-10]{18}(832,323)(844,323)(856,323)(868,323)(880,323)(892,323)(904,323)(916,323)(928,323)(940,323)(952,323)(964,323)(976,323)(988,323)(1000,323)(1012,323)(1024,323)(1036,323)(1048,323)(1060,323)(1072,323)(1084,323)(1096,323)(1108,323)(1120,323)(1132,323)(1145,323)(1157,323)(1169,323)(1181,323)(1193,323)(1205,323)(1217,323)(1229,323)(1241,323)(1253,323)(1265,323)(1277,323)(1289,323)(1301,323)(1313,323)(1325,323)(1337,323)(1349,323)(1361,323)(1373,323)(1385,323)(1397,323)(1409,323)(1421,323)(1433,323)
\end{picture}
\caption{$\Delta_S$ plotted versus $\delta_{ee}$ using Eq.(\ref{ndiagR}) for a fixed $\Delta_a = 3 \times 10^{-3} eV^2$.}
\end{figure}

In the light of these remarks we now consider briefly 
two extensions which introduce more freedom and can accommodate all the
present data: (1) More than one independent 
Higgs doublet coupling to charged leptons,
enabling off-diagonal flavor vertices for 
those doublets without a vacuum value\cite{BGS}. 
(2) Addition of a singlet doubly-charged scalar\cite{ZEEP,BABU,R}.

The neutrino Majorana mass in extended model (1) arises from the
graph in Fig. 5.

\begin{center}
\begin{figure}
\begin{picture}(400,200)(0,0)
\SetWidth{1.2}
{\LARGE
\ArrowLine(50,20)(120,20)
\ArrowLine(200,20)(120,20)
\ArrowLine(280,20)(200,20)
\ArrowLine(350,20)(280,20)
\DashArrowArcn(200,20)(80,180,90){5}
\DashArrowArcn(200,20)(80,90,0){5}
\Text(50,16)[rt]{$\nu_L$}
\Text(360,16)[rt]{$\nu_L$}
\Text(125,16)[rt]{$f_{ij}$}
\Vertex(120,20){3}
\Text(160,16)[t]{$l_L$}
\Text(240,16)[t]{$l_R$}
\Text(200,20)[]{$\times $}
\Text(200,30)[]{$m_l $}
\Text(200,100)[]{$\times $}
\Text(205,110)[b]{$\mu <v_i>$ }
\Text(130,70)[b]{$h^-$ }
\Text(280,70)[b]{$H^-_j$ }}
\end{picture}\\
\caption{Majorana neutrino mass for the minimal Zee model or for
the extended model (1)}
\end{figure}
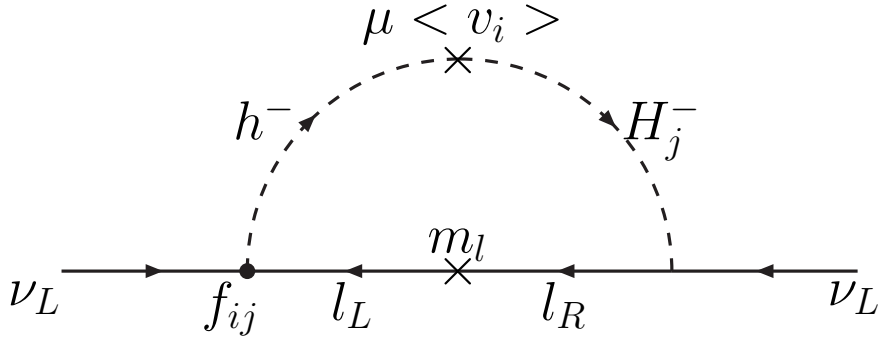
\end{center}


\begin{center}
\begin{figure}
\begin{picture}(400,200)(0,0)
\SetWidth{1.2}
{\LARGE
\ArrowLine(50,20)(120,20)
\ArrowLine(120,20)(280,20)
\ArrowLine(280,20)(350,20)
\DashArrowArcn(200,20)(80,180,90){5}
\DashArrowArcn(200,20)(80,90,0){5}
\Photon(200,100)(200,140){5}{5}
\Text(50,16)[rt]{$\mu$}
\Text(360,16)[rt]{$e$}
\Text(200,16)[t]{$\nu$}
\Text(180,110)[b]{$\gamma$ }
\Text(130,70)[b]{$H^-_i$ }
\Text(280,70)[b]{$H^-_i$ }}
\end{picture}\\
\caption{$\mu\rightarrow e \gamma $ in extended model (1)}
\end{figure}
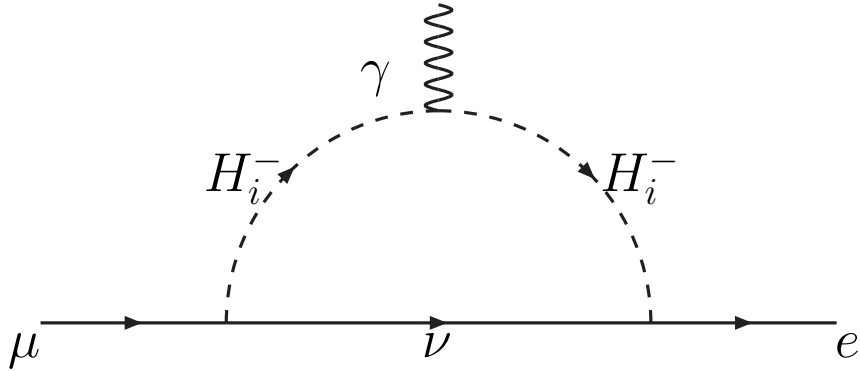
\end{center}

\newpage 
The Yukawa couplings in this model are 
\bea
g_1^{ij}\overline{\Psi^i_L}\phi_1 l_R^j 
+ g_2^{ij} \overline{\Psi^i_L}\phi_2 l_R^j +h.c. 
\eea
where $\Psi^i_L$ is left-handed lepton doublet and $l_R$ is right-handed
charged lepton. The $\phi_i$ are Higgs doublets. After SU(2) symmetry breaking 
the remaining charged Higgs coupling is 
\bea
\overline{\nu_L}^i \left[ \sqrt{2}\frac{v_2}{v_1} m^i \delta^{ij}
          - \frac{v_1^2+v_2^2}{v_1}g_2^{ij}\right] 
 l_R^j \frac{1}{ \sqrt{v_1^2+v_2^2} } (sin\theta H_1^+ - cos\theta H_2^+ )
\eea 
where $\theta$ is mixing angle between the charged Higgs and 
the singlet Zee scalar, $m^i$ is
the mass of charged lepton, $v_i$ are  the vacuum expectation 
values for the Higgs bosons and 
$H_i^+$ are mass eigenstates. The $g_2$ has flavor off-diagonal elements 
and by this feature diagonal neutrino masses arise from Fig. 5.   

The off-diagonal couplings also contain a $\mu\rightarrow e \gamma $ 
flavor violating decay\cite{petcov} arising from the diagram of Fig. 6. 
The branching ratio
\footnote{We neglected the contribution from the first term in Eq.(44) 
in this discussion because it 
is already smaller than the $g_2$ term if $g_2 \sim O(10^{-3})$.}
is 
\bea
BR(\mu\to e \gamma ) \propto \frac{\alpha}{48\pi G_F^2}
     \frac{(v_1^2+v_2^2)}{v_1^2}|g_2^{X\mu}g_2^{eX}|^2 
    \left[\frac{sin^2\theta}{M_1^2} + \frac{cos^2\theta}{M_2^2}\right]^2,
\eea
where $g_2$ is the Yukawa coupling with flavor off-diagonal elements. 
The experimental bound is $BR(\mu \to e \gamma) < 1.2 \times 10^{-11} $. From this bound, 
\bea
\frac{|g_2^{X\mu}g_2^{eX}|^2}{\bar{M}^4} < 2\times 10^{-17} (GeV)^{-4}, 
\eea
where $\frac{1}{\bar{M}^2} 
= \frac{sin^2\theta}{M_1^2} + \frac{cos^2\theta}{M_2^2}$. 
If we take $\bar{M} \sim 100GeV$, this implies 
\bea
g_2 < (4\sim7)\times 10^{-3}. 
\eea

While, the mass elements are 
\bea
\delta_{ee} m_1 &=& 
  M_{ee} \propto \frac{v_1^2 + v_2^2}{v_1} \mu g_2 f_{e\mu} m_\mu,\\
\frac{m_1}{\sqrt{2}} &\sim& M_{e\mu} 
\propto \frac{v_2}{v_1} \mu f_{e\mu} m_\mu^2. 
\eea
where $f_{e\mu}$ is the Zee scalar coupling 
and we neglected the term of 
$f_{e\tau}m_\tau$ because $f_{e\tau}/f_{e\nu} \propto m_\mu^2/m_\tau^2$ from 
$|M_{e\mu}| \sim |M_{e\tau}|$. 
The ratio is
\bea
\sqrt{2}\delta_{ee} \sim \frac{M_{ee}}{M_{e\mu}} &\propto& 
     \sim \frac{v_1^2+v_2^2}{v_2 m_\mu} g_2.
\eea

We need this ratio to be near to $2
\sqrt{2} \epsilon_3$. 
So we find: 
\bea
g_2 \sim 2 \sqrt{2} 
   \epsilon_3 m_\mu \frac{v_2}{v_1^2+v_2^2} \sim O(10^{-4}) .
\eea
estimated for the case $v_1 = v_2$. This condition will then comply with the bound from $\mu\rightarrow e\gamma $.   
Then extended model (1) is consistent with all the data.

\bigskip
\bigskip

Finally we consider extended model (2) which contains new couplings with a doubly-charged
scalar singlet $k^{++}$:

\bea
h_{XX} l_R^X CL_R^X k^{++} + h.c. + \kappa(h^+h^+k^{--} +h^-h^-k^{++}),
\eea
where $k^{++}$ is a doubly charged $SU(2)$ singlet scalar which couples to 
only right handed charged leptons. The contribution to 
the Majorana neutrino mass now comes from the two loop diagram of Fig.7. 

\begin{center}
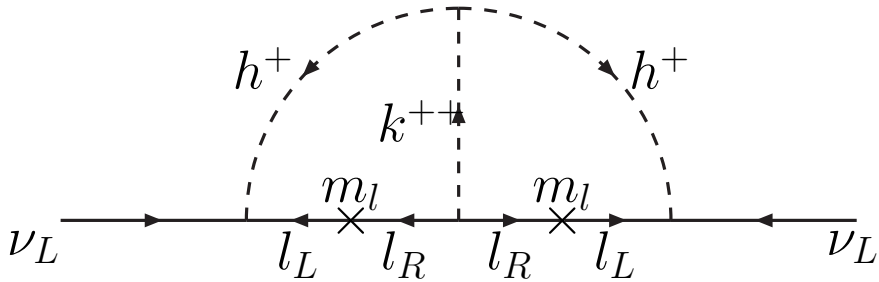
\begin{figure}
\begin{picture}(450,140)(0,0)
\SetWidth{1.2}
{\LARGE
\ArrowLine(50,20)(120,20)
\ArrowLine(160,20)(120,20)
\ArrowLine(200,20)(160,20)
\ArrowLine(200,20)(240,20)
\ArrowLine(240,20)(280,20)
\ArrowLine(350,20)(280,20)
\DashArrowArc(200,20)(80,90,180){5}
\DashArrowArcn(200,20)(80,90,0){5}
\DashArrowLine(200,20)(200,100){5}
\Text(50,16)[rt]{$\nu_L$}
\Text(360,16)[rt]{$\nu_L$}
\Text(140,16)[t]{$l_L$}
\Text(180,16)[t]{$l_R$}
\Text(220,16)[t]{$l_R$}
\Text(260,16)[t]{$l_L$}
\Text(160,20)[]{$\times $}
\Text(160,30)[]{$m_l $}
\Text(240,20)[]{$\times $}
\Text(240,30)[]{$m_l $}
\Text(190,50)[b]{$k^{++}$ }
\Text(130,70)[b]{$h^+$ }
\Text(280,70)[b]{$h^+$ }}
\end{picture}\\
\caption{Majorana neutrino mass from two loop diagram 
in the extended Model (2). }
\end{figure}
\end{center}

The diagonal mass element is 
\bea
M_{ee} &\propto& \frac{\kappa}{(4\pi)^4} 
  [h_{\mu\mu} \frac{m_\mu^2}{m_k^2} f^2_{e\mu}
   + h_{\tau\tau} \frac{m_\tau^2}{m_k^2} f^2_{e \tau} ] F \nn \\
       &\sim & \frac{\kappa}{(4\pi)^4} 
            h_{\mu\mu} \frac{m_\mu^2}{m_k^2} f^2_{e\mu}, 
\eea
where $m_X$ ($X=\mu,\tau $) is the charged lepton mass, $m_k$ is the mass 
of doubly charged scalar and $F$ show some log function. 
Here we neglected 
the term $f_{e\tau}^2$ for the same reason we neglected it in Eq.(48). 
To get the required mass element, $M_{ee}$ has to be  around 
$2 \epsilon_3 m_1$. 
Namely, the ratio between the diagonal and off-diagonal elements 
of the neutrino mass term should be near to 2$\epsilon_3$. 
\bea
\frac{M_{ee}}{M_{e\mu}} \sim \frac{1}{16 \pi^2} \frac{\kappa }{\mu}
              \frac{v_1}{v_2}\frac{m_h^2}{m_k^2}h_{\mu\mu}f_{e\mu} 
\eea
We may always realize this condition which depends on unknown parameters,
and hence the extended model (2) is consistent with all the neutrino data. 

\bigskip
\bigskip

\section{Discussion}
The minimal Zee model is very economical as a simple way to introduce
neutrino mass into the standard model. However, it is seen to be barely
consistent with the combination
of SuperKamiokande and SNO data. It would need $\Delta_S > \Delta_a$
which looks like a considerable stretch from the observations at hand.

On the other hand, if we enrich the model by either (1) adding
further Higgs
doublets coupling to the charged leptons and to the singlet
charged scalar or (2) by adding a
doubly-charged singlet scalar, there is enough freedom to accommodate 
the SNO data.

\bigskip
\bigskip

{\it Acknowledgments.}
This work was supported in part by the US Department of Energy
under Grant No. DE-FG02-97ER-41036.

\end{document}